\documentclass[11pt]{article}
\usepackage{graphicx}
\usepackage{epsfig}
\usepackage{amsmath,amssymb}

\newcommand{\cn}{\mathrm{cn}}

\newcommand{\sech}{\rm sech}
\newcommand{\csch}{\rm csch}
\begin{document}

\markboth{Sirendaoreji}
{Weierstrass elliptic function solutions and their degenerated solutions of the mKdV equation}

\title{A method for constructing Weierstrass elliptic function
solutions and their degenerated solutions of the mKdV equation}

\author{Sirendaoreji\footnote{E-mail: siren@imnu.edu.cn}\\{\small
       Mathematical Science College, Inner Mongolia Normal University,}\\
       {\small Huhhot 010022, Inner Mongolia, P. R. China}}
\date{}
\maketitle

\begin{abstract}
A Weierstrass type projective Riccati equation expansion method
is proposed by using the Weierstrass elliptic function solutions
of the projective Riccati equations and the conversion formulas
which transform the Weierstrass elliptic functions into the
hyperbolic and the trigonometric functions.
The Weierstrass elliptic function solutions and their degenerated
solutions including the solitary wave and the periodic wave solutions
of the mKdV equation are obtained by using the proposed method.
\end{abstract}

\section{Introduction}
\label{sec-1}
Traveling wave solutions of nonlinear evolution equations (NLEEs) can help
people to understand and explain nonlinear phenomena in many areas of applied
science, such as plasma physics, optical fiber, fluid mechanics, biology and
chemistry, etc. Therefore, the problem of how to find new methods for
seeking traveling wave solutions of NLEEs has become a hot topic in soliton
theory. In the past few decades, various direct methods, such as the
tanh--function method\cite{1}, the auxiliary equation method\cite{2},
the Riccati equation expansion method\cite{3}, the unified Riccati equation
expansion method\cite{4}, the Jacobi elliptic function expansion method\cite{5},
the Weierstrass elliptic function method\cite{6,7,8,9,10}, the projective
Riccati equation expansion method\cite{11,12,13,14,15}, and others have
been proposed to find exact traveling wave solutions of NLEEs.\par
Among them, the projective Riccati equation expansion method usually can give
some new traveling wave solutions of NLEEs which are different from those solutions
obtained by other direct methods.
In Ref.\cite{16}, two groups of Weierstrass elliptic function solutions of the
projective Riccati equations were used to construct the Weierstrass
elliptic function solutions of the Burgers equation and a nonlinear
dispersive--dissipative equation.
Also in Refs.\cite{17--18}, the Weierstrass elliptic function solutions
of the modified Korteweg--de Vries (mKdV) equation were obtained by using the projective
Riccati equation expansion method. Especially, in Ref.\cite{17} the author attempt
to transform the Weierstrass elliptic function solutions of the mKdV equation into the
hyperbolic and trigonometric function solutions using the following conversion formula
\begin{align}
\wp(z,g_2,g_2)=e_2-(e_2-e_3)\,\cn^2\left(\sqrt{e_1-e_3}z,m\right),
m=\sqrt{\frac {e_2-e_3}{e_1-e_3}},
\label{eq1:1}
\end{align}
which depends on the roots $e_1,e_2,e_3~(e_1\geqslant{e_2}\geqslant{e_3})$
of the third order polynomial equation $P(w)\equiv{w^3-g_2w-g_3}=0$. It is
no doubt that the formula (\ref{eq1:1}) can be used to convert the Weierstrass
elliptic function solutions of NLEEs into the Jacob elliptic function solutions,
and then the hyperbolic and trigonometric function solutions of NLEEs can be
obtained by taking the limit of modulus $m\rightarrow{1}$ and $m\rightarrow{0}$.
But it is already pointed out in~\cite{19,20,21} that the
solutions obtained by this process may be incorrect.\par
In order to overcome this difficulty, in Ref.~\cite{21} we have constructed
four new conversion formulas which can directly convert the Weierstrass
elliptic function solutions of NLEEs into the hyperbolic and trigonometric
function solutions. More importantly, these formulas do not depend on the
roots of the above third order polynomial equation, not need to use the
Jacobi elliptic functions in the converting process, and can ensure
that the converted solutions of NLEEs are correct.\par
This indicates that if we can find some new Weierstrass elliptic function solutions
of the projected Riccati equations, then we can use these Weierstrass elliptic
function solutions and our conversion formulas to give some new solitary
wave and trigonometric periodic wave solutions of NLEEs.
Therefore, this paper aims to give more Weierstrass elliptic function solutions of
the projective Riccati equations and use our conversion formulas to propose a direct
method which we called the Weierstrass type projective Riccati equation expansion
method. Finally, we shall take the mKdV equation as an
illustrative example to show the effectiveness of our method.\par
This paper is organized as follows. In the next section, four groups of Weierstrass
elliptic function solutions of the projective Riccati equations are constructed. And
by using these Weierstrass elliptic function solutions and our conversion formulas,
the called Weierstrass type projective Riccati equation expansion method is proposed
to find exact traveling wave solutions to NLEEs.
In Sec.~\ref{sec-3}, the proposed method is applied to construct the Weierstrass
elliptic function solutions, the solitary and periodic wave solutions of the mKdV
equation. The conclusions are given in Sec.~\ref{sec-4}.
\section{Weierstrass type projective Riccati equation expansion method}
\label{sec-2}
The Weierstrass elliptic function $w=\wp(\xi,g_2,g_3)$ is defined as the inverse
function of the Weierstrass elliptic integral\cite{22,23}
\begin{align}
\xi=\int_\infty^w {\frac {dt}{\sqrt{4t^3-g_2t-g_3}}},
\label{eq2:1}
\end{align}
or the solution of the following nonlinear ordinary differential equation (ODE)
\begin{align}
\label{eq2:2}
\left({\frac {dw}{d\xi}}\right)^2=P(w)=4w^3-g_2w-g_3,
\end{align}
where the real parameters $g_2,g_3$ are called invariants.\par
In the following we shall consider the projective Riccati equations of the form
\begin{align}
\left\{\begin{aligned}
&F^{\prime}(\xi)=pF(\xi)G(\xi),\\
&G^{\prime}(\xi)=q+pG^2(\xi)-rF(\xi),
\end{aligned}\right.
\label{eq2:3}
\end{align}
where $F$,$G$ are unknown functions of the variable $\xi$ and $p,q,r$
are constants. \par
By using the direct assumption approach we can construct
the Weierstrass elliptic function solutions of Eqs.~(\ref{eq2:3}) as following
\begin{align}
&\left\{\begin{aligned}
   &F(\xi)={\frac q{6r}}+{\frac 2{pr}}\wp(\xi,g_2,g_3),\\
   &G(\xi)={\frac {12\wp^\prime(\xi,g_2,g_3)}{p\left[pq+12\wp(\xi,g_2,g_3)\right]}},
   \end{aligned}\right.
   \label{eq2:4}\\
&{\qquad}G^2(\xi)=-{\frac qp}+{\frac {2r}{p}}F(\xi),
\label{eq2:5}\\
&\left\{\begin{aligned}
&F(\xi)={\frac {5q}{6r}}+{\frac {5q^2p}{72r\wp(\xi,g_2,g_3)}},\\
&G(\xi)={\frac {-q\wp^\prime(\xi,g_2,g_3)}{\left[qp+12\wp(\xi,g_2,g_3)\right]\wp(\xi,g_2,g_3)}},
\end{aligned}\right.
\label{eq2:6}\\
&{\qquad}G^2(\xi)=-{\frac qp}+{\frac {2r}{p}}F(\xi)-{\frac {24r^2}{25pq}}F^2(\xi),
\label{eq2:7}\\
&\left\{\begin{aligned}
&F(\xi)={\frac {(2+q)\left[pq+12\wp(\xi,g_2,g_3)\right]}{r\left[12p+pq+12\wp(\xi,g_2,g_3)\right]}},\\
&G(\xi)={\frac {\wp^\prime(\xi,g_2,g_3)}{\left(\wp(\xi,g_2,g_3)+{\frac p2}
+{\frac {pq}{12}}\right)^2-{\frac {p^2}{4}}}},
\end{aligned}\right.
\label{eq2:8}\\
&{\qquad}G^2(\xi)=-{\frac qp}+{\frac {2r}{p}}F(\xi)-{\frac {r^2(q+4)}{p(q+2)^2}}F^2(\xi),
\label{eq2:9}\\
&\left\{\begin{aligned}
&F(\xi)={\frac {q(p^2+2)\left[pq+12\wp(\xi,g_2,g_3)\right]}{pr\left[12q+p^2q+12p\wp(\xi,g_2,g_3)\right]}},\\
&G(\xi)={\frac {q\wp^\prime(\xi,g_2,g_3)}{\left(p\wp(\xi,g_2,g_3)+{\frac q2}
+{\frac {p^2q}{12}}\right)^2-{\frac {q^2}{4}}}},
\end{aligned}\right.
\label{eq2:10}\\
&{\qquad}G^2(\xi)=-{\frac qp}+{\frac {2r}{p}}F(\xi)-{\frac {pr^2(p^2+4)}{q(p+2)^2}}F^2(\xi),
\label{eq2:11}
\end{align}
where the invariants $g_2$ and $g_3$ are given by
\begin{align}
g_2={\frac {p^2q^2}{12}},g_3={\frac {p^3q^3}{216}}.
\label{eq2:12}
\end{align}
Here the solutions (\ref{eq2:4}) and (\ref{eq2:6}) are previously known\cite{16,17,18},
but the solutions (\ref{eq2:8}) and (\ref{eq2:10}) are new.\par
We find that the Weierstrass elliptic function can degenerate to the hyperbolic and
trigonometric function by using the following conversion formulas\cite{21}
\begin{align}
&\wp(\xi,{\frac {\theta^2}{12}},-{\frac {\theta^3}{216}})
={\frac {\theta}{12}}-{\frac {\theta}{4}}\sech^2({\frac {\sqrt{\theta}}{2}}\,\xi),\theta>0,
\label{eq2:13}\\
&\wp(\xi,{\frac {\theta^2}{12}},-{\frac {\theta^3}{216}})
={\frac {\theta}{12}}+{\frac {\theta}{4}}\csch^2({\frac {\sqrt{\theta}}{2}}\,\xi),\theta>0,
\label{eq2:14}\\
&\wp(\xi,{\frac {\theta^2}{12}},-{\frac {\theta^3}{216}})
={\frac {\theta}{12}}-{\frac {\theta}{4}}\sec^2({\frac {\sqrt{-\theta}}{2}}\,\xi),\theta<0,
\label{eq2:15}\\
&\wp(\xi,{\frac {\theta^2}{12}},-{\frac {\theta^3}{216}})
={\frac {\theta}{12}}-{\frac {\theta}{4}}\csc^2({\frac {\sqrt{-\theta}}{2}}\,\xi),\theta<0,
\label{eq2:16}
\end{align}
where $\theta$ is a real number.\par
A given NLEE with respect to two variables $x$ and $t$ is of the form
\begin{align}
P(u,u_x,u_t,u_{xx},u_{xt},u_{tt},u_{xxx},\cdots)=0,
\label{eq2:17}
\end{align}
where the subscripts denote the partial derivatives,$P$ is a polynomial in unknown
function $u(x,t)$ and its derivatives.\par
The Weierstrass type projective Riccati equation expansion method proposed
here follows the following five steps.\par
{\em Step 1:} Making the wave transformation
\begin{align}
u(x,t)=u(\xi),\xi=x-{\omega}t,
\label{eq2:18}
\end{align}
we may exchange the Eq.~(\ref{eq2:17}) into the following ODE
\begin{align}
H(u,u^\prime,u^{\prime\prime},\cdots)=0,
\label{eq2:19}
\end{align}
where primes denote the derivatives with respect to $\xi$ and the
wave speed $\omega$ is a constant to be determined later.\par
{\em Step 2:} Assume that the Eq.~(\ref{eq2:19}) has the truncated formal
series solution
\begin{align}
u(\xi)=a_0+\sum_{i=1}^n F^{i-1}(\xi)\left(a_iF(\xi)+b_iG(\xi)\right),
\label{eq2:20}
\end{align}
where $F(\xi)$ and $G(\xi)$ are the Weierstrass elliptic function solution
of the projective Riccati Eqs.~(\ref{eq2:3}), $n$ is an integer number which
can be determined by balancing the highest order derivative terms with the
highest power nonlinear terms in (\ref{eq2:19}),$a_0$,$a_i,b_i~(i=1,2,\cdots,n)$
are undetermined constants and $a_n^2+b_n^2\not=0$.
\par
{\em Step 3:} Substituting (\ref{eq2:20}), (\ref{eq2:3}) together with one
of the relations (\ref{eq2:5}), (\ref{eq2:7}), (\ref{eq2:9}), (\ref{eq2:11})
into (\ref{eq2:19}) and equating the coefficients of like
powers of $F^i(\xi)G^j(\xi)$ to zero yields a set of algebraic equations.
Solving this set of algebraic equations with the aid of Maple or Mathematica
we can determine the values of $\omega,a_0$ and $a_i,b_i~(i=1,2,\cdots,n)$.\par
{\em Step 4:} Putting each solutions of the algebraic equations obtained in
{\em Step 3} together with the Weierstrass elliptic function solutions $F(\xi),G(\xi)$
of the projective Riccati equations into (\ref{eq2:20}) and using (\ref{eq2:18}),
we obtain the Weierstrass type traveling wave solutions of the Eq.~(\ref{eq2:17}).\par
{\em Step 5:} Taking the conversion formulas (\ref{eq2:13})--(\ref{eq2:16})
into the Weierstrass elliptic function type traveling wave solutions obtained
in {\em Step 4}, we get the solitary wave and the periodic wave solutions of
the Eq.~(\ref{eq2:17}).
\section{Solutions of the mKdV equation}
\label{sec-3}
Now let us consider the mKdV equation
\begin{align}
\label{eq3:1}
u_t+{\alpha}u^2u_x+{\beta}u_{xxx}=0,
\end{align}
where $\alpha,\beta$ are constants.\par
Taking the wave transformation (\ref{eq2:18}) into (\ref{eq3:1})
we get the following ODE
\begin{align}
\label{eq3:2}
-{\omega}u^\prime(\xi)+{\alpha}u^2(\xi)u^\prime(\xi)+{\beta}u^{\prime\prime\prime}(\xi)=0.
\end{align}
By using the homogenous balance method we can determine that $n=1$.Thus
the solution of Eq. (\ref{eq3:2}) can be chosen as
\begin{align}
u(\xi)=a_0+a_1F(\xi)+b_1G(\xi),
\label{eq3:3}
\end{align}
where $F(\xi),G(\xi)$ are the solution of the projective Riccati equations
(\ref{eq2:3}),$a_0,a_1,b_1$ are undermined constants.\par
(1)\;Taking (\ref{eq3:3}) with (\ref{eq2:3}),(\ref{eq2:5}) into (\ref{eq3:2})
and setting the coefficients of $F^iG^j~(i=1,2,3;j=0,1)$ to zero we
obtain a set of algebraic equations
\begin{align*}
\left\{\begin{aligned}
&{\alpha}a_1^3p=0,5{\alpha}a_1^2b_1r=0,\\
&6{\alpha}a_1a_0b_1r-2{\alpha}a_1^2b_1q
    +{\frac {2{\alpha}b_1^3r^2}{p}}+3{\beta}b_1r^2p=0,\\
&-2{\alpha}a_1a_0b_1q-{\frac {{\alpha}b_1^3qr}{p}}-{\beta}b_1rpq
    +{\alpha}a_0^2b_1r-{\omega}b_1r=0,\\
&2{\alpha}a_0a_1^2p+4{\alpha}a_1b_1^2r+6{\beta}a_1p^2r=0,\\
&{\alpha}a_0^2a_1p+2{\alpha}a_0b_1^2r-{\alpha}a_1b_1^2q-{\beta}a_1p^2q-{\omega}a_1p=0.
\end{aligned}\right.
\end{align*}
Solving this set of algebraic equations with use of Maple we get
\begin{align}
a_0=0,a_1=0,b_1=\varepsilon{p}\sqrt{-\frac {3\beta}{2\alpha}},\omega={\frac {\beta{pq}}{2}},\varepsilon=\pm{1}.
\label{eq3:4}
\end{align}
Substituting (\ref{eq3:4}),(\ref{eq2:4}),(\ref{eq2:18}) into (\ref{eq3:3}) we
obtain the Weierstrass type traveling wave solution of the mKdV equation
\begin{align}
u(x,t)={\frac {12\varepsilon\sqrt{-\frac {3\beta}{2\alpha}}\wp^\prime(x-{\frac {\beta{pq}}{2}}t,g_2,g_3)}{pq+12\wp(x-{\frac {\beta{pq}}{2}}t,g_2,g_3)}},
\label{eq3:5}
\end{align}
where the invariants $g_2,g_3$ are determined by (\ref{eq2:12}).\par
On substituting the conversion formulas (\ref{eq2:13})--(\ref{eq2:16}) with
$\theta=-qp$ into Eq.~(\ref{eq3:5}), we find that the Weierstrass elliptic
function solution (\ref{eq3:5}) degenerate to the following solitary wave
and periodic wave solutions of the mKdV equation
\begin{align*}
&u_1^{(1)}(x,t)=-{\frac {\varepsilon}{2}}\sqrt{\frac {6\beta{pq}}{\alpha}}
    \tanh{\frac {\sqrt{-pq}}{2}}(x-{\frac {\beta{pq}}{2}}t),
    pq<0,\alpha\beta<0,\\
&u_2^{(1)}(x,t)=-{\frac {\varepsilon}{2}}\sqrt{\frac {6\beta{pq}}{\alpha}}
    \coth{\frac {\sqrt{-pq}}{2}}(x-{\frac {\beta{pq}}{2}}t),
    pq<0,\alpha\beta<0,
\end{align*}
\begin{align*}
&u_3^{(1)}(x,t)={\frac {\varepsilon}{2}}\sqrt{-\frac {6\beta{pq}}{\alpha}}
    \tan{\frac {\sqrt{pq}}{2}}(x-{\frac {\beta{pq}}{2}}t),
    pq>0,\alpha\beta<0,\\
&u_4^{(1)}(x,t)=-{\frac {\varepsilon}{2}}\sqrt{-\frac {6\beta{pq}}{\alpha}}
    \cot{\frac {\sqrt{pq}}{2}}(x-{\frac {\beta{pq}}{2}}t),
    pq>0,\alpha\beta<0,
\end{align*}
where $p,q$ are free parameters.\par
(2)\;Taking (\ref{eq3:3}) with (\ref{eq2:3}), (\ref{eq2:7}) into (\ref{eq3:2})
and setting the coefficients of $F^iG^j~(i=1,2,3,4,5;j=0,1)$ to zero
leads the following set of algebraic equations
\begin{align*}
\left\{\begin{aligned}
&{\alpha}a_1^3p-{\frac {144{\beta}a_1p^2r^2+72{\alpha}b_1^2a_1r^2}{25q}}=0,\\
&{\frac {3456{\beta}b_1pr^4}{625q^2}}-{\frac {72{\alpha}a_1^2b_1r^2}{25q}}
    +{\frac {576{\alpha}b_1^3r^4}{625q^2p}}=0,\\
&4{\alpha}a_1b_1^2r+6{\beta}a_1p^2r+2{\alpha}a_1^2a_0p-{\frac {48{\alpha}b_1^2a_0r^2}{25q}}=0,\\
&5{\alpha}a_1^2b_1r-{\frac {72{\alpha}b_1^3r^3}{25qp}}
    -{\frac {288{\beta}b_1pr^3+96{\alpha}b_1a_0a_1r^2}{25q}}=0,\\
&-2{\alpha}a_1a_0b_1q-{\frac {{\alpha}b_1^3rq}{p}}-{\beta}b_1rpq
    +{\alpha}a_0^2b_1r-{\omega}b_1r=0,\\
&6{\alpha}a_1a_0b_1r+{\frac {24{\omega}b_1r^2-24{\alpha}a_0^2b_1r^2}{25q}}
    -2{\alpha}a_1^2b_1q+{\frac {74{\alpha}b_1^3r^2}{25p}}+{\frac {171{\beta}b_1pr^2}{25}}=0,\\
&{\alpha}a_0^2a_1p+2{\alpha}a_0b_1^2r-{\alpha}a_1b_1^2q-{\beta}a_1p^2q-{\omega}a_1p=0.
\end{aligned}\right.
\end{align*}
Using Maple to solve the above algebraic equations we obtain
\begin{align}
a_0={\frac {5\varepsilon\sqrt{\alpha\omega}}{3\alpha}},b_1=0,
q={\frac {16\omega}{9p\beta}},r=-{\frac {5a_1\varepsilon\sqrt{\alpha\omega}}{9p\beta}},
\varepsilon=\pm{1},
\label{eq3:6}
\end{align}
where $\alpha\omega>0$ and $p$ is a free parameter.\par
Inserting (\ref{eq3:6}), (\ref{eq2:6}) and (\ref{eq2:18}) into (\ref{eq3:3})
we obtain the following Weierstrass type traveling wave solution of the mKdV equation
\begin{align}
u(x,t)=-{\frac {\varepsilon\omega\left(32\omega+81\beta\wp(x-{\omega}t,g_2,g_3)\right)}
     {81\beta\sqrt{\alpha\omega}\wp(x-{\omega}t,g_2,g_3)}},
\label{eq3:7}
\end{align}
where $\alpha\omega>0$,the invariants $g_2$,$g_3$
are given by
\begin{align}
\label{eq3:8}
g_2={\frac {64\omega^2}{243\beta^2}},g_2={\frac {512\omega^3}{19683\beta^3}}.
\end{align}
Substituting the conversion formulas (\ref{eq2:13})--(\ref{eq2:16}) with
$\theta=-qp=-{\frac {16\omega}{9\beta}}$ into (\ref{eq3:7}) and using (\ref{eq3:8})
we obtain the solitary wave like solutions and the periodic wave solutions of
the mKdV equation as following
\begin{align*}
&u_1^{(2)}(x,t)={\frac {\varepsilon\omega\left(5\cosh^2{\frac 23}
    \sqrt{-\frac {\omega}{\beta}}(x-{\omega}t)+9\right)}
    {3\sqrt{\alpha\omega}\left(\cosh^2{\frac 23}
    \sqrt{-\frac {\omega}{\beta}}(x-{\omega}t)-3\right)}},
    \alpha\omega>0,\beta\omega<0,
\end{align*}
\begin{align*}
&u_2^{(2)}(x,t)={\frac {\varepsilon\omega\left(5\sinh^2{\frac 23}
    \sqrt{-\frac {\omega}{\beta}}(x-{\omega}t)-9\right)}
    {3\sqrt{\alpha\omega}\left(\sinh^2{\frac 23}
    \sqrt{-\frac {\omega}{\beta}}(x-{\omega}t)+3\right)}},
    \alpha\omega>0,\beta\omega<0,\\
&u_3^{(2)}(x,t)={\frac {\varepsilon\omega\left(5\cos^2{\frac 23}
    \sqrt{\frac {\omega}{\beta}}(x-{\omega}t)+9\right)}
    {3\sqrt{\alpha\omega}\left(\cos^2{\frac 23}
    \sqrt{\frac {\omega}{\beta}}(x-{\omega}t)-3\right)}},
    \alpha\omega>0,\beta\omega>0,\\
&u_4^{(2)}(x,t)={\frac {\varepsilon\omega\left(5\sin^2{\frac 23}
    \sqrt{\frac {\omega}{\beta}}(x-{\omega}t)+9\right)}
    {3\sqrt{\alpha\omega}\left(\sin^2{\frac 23}
    \sqrt{\frac {\omega}{\beta}}(x-{\omega}t)-3\right)}},
    \alpha\omega>0,\beta\omega>0,
\end{align*}
where $\omega$ is a free parameter.\par
(3)\; Inserting (\ref{eq3:3}) with (\ref{eq2:3}), (\ref{eq2:9}) into (\ref{eq3:2})
and setting the coefficients of $F^iG^j~(i=1,2,3,4;j=0,1)$ to zero
we obtain the following set of algebraic equations
\begin{align*}
\left\{\begin{aligned}
&{\alpha}a_1^3p-{\frac {24{\beta}a_1p^2r^2+3{\alpha}b_1^2a_1r^2q
    +6{\beta}a_1p^2r^2q+12{\alpha}b_1^2a_1r^2}{(q+2)^2}}=0,\\
&{\alpha}a_0^2b_1r-{\omega}b_1r-2{\alpha}a_1a_0b_1q
    -{\frac {{\alpha}b_1^3rq}{p}}-2{\beta}b_1prq=0,\\
&4{\alpha}a_1b_1^2r+2{\alpha}a_1^2a_0p+6{\beta}a_1p^2r
    -{\frac {2{\alpha}b_1^2a_0r^2q+8{\alpha}b_1^2a_0r^2}{(q+2)^2}}=0,\\
&5{\alpha}a_1^2b_1r-{\frac {14{\beta}b_1pr^3q+4{\alpha}b_1a_0a_1r^2q
    +16{\alpha}b_1a_0a_1r^2+56{\beta}b_1pr^3}{(q+2)^2}}\\
    &{\quad}-{\frac {3{\alpha}b_1^3r^3q+12{\alpha}b_1^3r^3}{p(q+2)^2}}=0,\\
&{\frac {6{\beta}b_1pr^4q^2+48{\beta}b_1pr^4q+96{\beta}b_1pr^4}{(q+2)^4}}
     +{\frac {{\alpha}b_1^3r^4q^2+8{\alpha}b_1^3r^4q+16{\alpha}b_1^3r^4}{(q+2)^4p}}\\
     &{\quad}-{\frac {12{\alpha}a_1^2b_1r^2+3{\alpha}a_1^2b_1r^2q}{(q+2)^2}}=0,\\
&{\frac {{\alpha}b_1^3r^2q^2+4{\alpha}b_1^3r^2q}{p(q+2)^2}}
     +{\frac {2{\alpha}b_1^3r^2}{p}}-2{\alpha}a_1^2b_1q+6{\beta}b_1pr^2+6{\alpha}a_1a_0b_1r\\
     &{\quad}+{\frac {4{\beta}b_1pr^2q^2+16{\beta}b_1pr^2q-4{\alpha}a_0^2b_1r^2
     +{\omega}b_1r^2q-{\alpha}a_0^2b_1r^2q+4{\omega}b_1r^2}{(q+2)^2}}=0,\\
&{\alpha}a_0^2a_1p+2{\alpha}a_0b_1^2r-{\alpha}a_1b_1^2q-{\beta}a_1p^2q-{\omega}a_1p=0.
\end{aligned}\right.
\end{align*}
Solving this set of algebraic equations with aid of Maple we obtain
\begin{align}
&a_0=\varepsilon(q+2)\sqrt{\frac {3\omega}{\alpha(q^2+4q+12)}},
a_1=-{\frac {2\sqrt{3}\varepsilon\omega(q+4)r}{(q+2)\alpha
    \sqrt{\frac {\omega(q^2+4q+12)}{\alpha}}}},\nonumber\\
&{\qquad} b_1=0,p={\frac {2\omega(q+4)}{(q^2+4q+12)\beta}}.
\label{eq3:9}
\end{align}
From which we calculate that
\begin{align}
g_2={\frac {q^2\omega^2(q+4)^2}{3(q^2+4q+12)^2\beta^2}},
g_3={\frac {q^3\omega^3(q+4)^3}{27(q^2+4q+12)^3\beta^3}}.
\label{eq3:10}
\end{align}
Substituting (\ref{eq3:9}) with (\ref{eq3:10}),(\ref{eq2:8}) into (\ref{eq3:3})
we get the Weierstrass type traveling wave solution of the mKdV equation
\begin{align}
u(x,t)=-{\frac {\sqrt{3}\omega\varepsilon\left[6\beta(q+6)(q^2+4q+12)\wp(\xi,g_2,g_3)
      +\omega(q+4)(q^2-6q-24)\right]}{\sqrt{\alpha\omega(q^2+4q+12)}
      \left[6(q^2+4q+12)\beta\wp(\xi,g_2,g_3)+\omega(q+12)(q+4)\right]}},
\label{eq3:11}
\end{align}
where $\xi=x-{\omega}t$ and the invariants $g_2,g_3$ are given by (\ref{eq3:10}).\par
Taking the conversion formulas (\ref{eq2:13})--(\ref{eq2:16}) with
$\theta=-qp=-{\frac {2\omega(q+4)q}{(q^2+4q+12)\beta}}$ into (\ref{eq3:11})
leads the solitary and periodic wave solutions of the mKdV equation
\begin{align*}
&u_1^{(3)}(x,t)=\varepsilon\sqrt{\frac {3\omega}{\alpha(q^2+4q+12)}}
    \left(\frac {4(q+2)\cosh^2\eta-q(q+6)}{4\cosh^2\eta+q}\right),\\
&u_2^{(3)}(x,t)=-\varepsilon\sqrt{\frac {3\omega}{\alpha(q^2+4q+12)}}
    \left(\frac {4(q+2)\sinh^2\eta+q(q+6)}{4\sinh^2\eta-q}\right),\\
&{\quad}\eta={\frac 12}\sqrt{-\frac {2\omega(q+4)q}{(q^2+4q+12)\beta}}(x-{\omega}t),
        \alpha\omega>0,\beta\omega(q+4)q<0,\\
&u_3^{(3)}(x,t)=\varepsilon\sqrt{\frac {3\omega}{\alpha(q^2+4q+12)}}
    \left(\frac {4(q+2)\cos^2\zeta-q(q+6)}{4\cos^2\zeta+q}\right),\\
&u_4^{(3)}(x,t)=\varepsilon\sqrt{\frac {3\omega}{\alpha(q^2+4q+12)}}
    \left(\frac {4(q+2)\sin^2\zeta-q(q+6)}{4\sin^2\zeta+q}\right),\\
&{\quad}\zeta={\frac 12}\sqrt{\frac {2\omega(q+4)q}{(q^2+4q+12)\beta}}(x-{\omega}t),
       \alpha\omega>0,\beta\omega(q+4)q>0,
\end{align*}
where $q$ and $\omega$ are free parameters.\par
(4)\,Taking (\ref{eq3:3}) with (\ref{eq2:3}),(\ref{eq2:11}) into (\ref{eq3:2})
and setting the coefficients of $F^iG^j~(i=1,2,3,4;j=0,1)$ to zero
we obtain the following set of algebraic equations
\begin{align*}
\left\{\begin{aligned}
&{\alpha}a_0^2b_1r-{\omega}b_1r-2{\alpha}a_1a_0b_1q-{\frac {{\alpha}b_1^3rq}{p}}-{\beta}b_1rpq=0,\\
&2{\alpha}a_1^2a_0p-{\frac {2{\alpha}b_1^2a_0p^4r^2+8{\alpha}b_1^2a_0p^2r^2}{q(p^2+2)^2}}
    +4{\alpha}a_1b_1^2r+6{\beta}a_1p^2r=0,\\
&{\alpha}a_1^3p-{\frac {3{\alpha}b_1^2a_1p^4r^2+12{\alpha}b_1^2a_1p^2r^2
    +6{\beta}a_1p^6r^2+24{\beta}a_1p^4r^2}{q(p^2+2)^2}}=0,\\
&{\frac {3{\alpha}b_1^3p^3r^3+12{\alpha}b_1^3pr^3
    +12{\beta}b_1p^5r^3+48{\beta}b_1p^3r^3+4{\alpha}b_1a_0a_1p^4r^2+16{\alpha}b_1a_0a_1p^2r^2}{q(p^2+2)^2}}\\
    &{\quad}-5{\alpha}a_1^2b_1r=0,\\
&{\frac {6{\beta}b_1p^9r^4+48{\beta}b_1p^7r^4+96{\beta}b_1p^5r^4
    +8{\alpha}b_1^3p^5r^4+16{\alpha}b_1^3p^3r^4+{\alpha}b_1^3p^7r^4}{q^2(p^2+2)^4}}\\
    &{\quad}-{\frac {3{\alpha}a_1^2b_1p^4r^2+12{\alpha}a_1^2b_1p^2r^2}{q(p^2+2)^2}}=0,\\
&{\frac {{\omega}b_1p^4r^2+4{\omega}b_1p^2r^2-4{\alpha}a_0^2b_1p^2r^2-{\alpha}a_0^2b_1p^4r^2}{q(p^2+2)^2}}
     -2{\alpha}a_1^2b_1q+{\frac {2{\alpha}b_1^3r^2}{p}}+3{\beta}b_1r^2p\\
&{\quad}+6{\alpha}a_1a_0b_1r+{\frac {{\alpha}b_1^3p^3r^2+4{\alpha}b_1^3pr^2
     +4{\beta}b_1p^5r^2+16{\beta}b_1p^3r^2}{(p^2+2)^2}}=0,\\
&{\alpha}a_0^2a_1p+2{\alpha}a_0b_1^2r-{\alpha}a_1b_1^2q-{\beta}a_1p^2q-{\omega}a_1p=0.
\end{aligned}\right.
\end{align*}
Solving this set of algebraic equations with use of Maple we find that
\begin{align}
&a_0=0,b_1=p\sqrt{-\frac {3\beta}{2\alpha}},q={\frac {2\omega}{p\beta}},
r={\frac {2\varepsilon(p^2+2)a_1}{p^2\beta}}\sqrt{\frac {\alpha\omega}{3p^2+12}},
\label{eq3:12}\\
&a_0=0,b_1=-p\sqrt{-\frac {3\beta}{2\alpha}},q={\frac {2\omega}{p\beta}},
r={\frac {2\varepsilon(p^2+2)a_1}{p^2\beta}}\sqrt{\frac {\alpha\omega}{3p^2+12}},
\label{eq3:13}
\end{align}
where $\varepsilon=\pm{1}$.\par
From (\ref{eq3:12}) and (\ref{eq3:13}), the invariants $g_2$ and $g_3$ are
calculated to be
\begin{align}
g_2={\frac {\omega^2}{3\beta^2}},g_3={\frac {\omega^3}{27\beta^3}}.
\label{eq3:14}
\end{align}
Taking (\ref{eq3:12}) and (\ref{eq3:13}) with (\ref{eq3:14}),
(\ref{eq2:18}) into (\ref{eq3:3}),respectively,we obtain the Weierstrass type
traveling wave solution of the mKdV equation
\begin{align}
&u(x,t)=\varepsilon\omega\sqrt{\frac {3(p^2+4)}{\alpha\omega}}\left(
    {\frac {36\beta\varepsilon\sqrt{-\frac {2\beta\omega}{p^2+4}}\wp^\prime(\xi)
    +12\beta{p}\left(3\beta\wp(\xi)+\omega\right)\wp(\xi)+\omega^2p}
    {36\beta^2p^2\wp^2(\xi)+12\beta\omega(p^2+6)\wp(\xi)+\omega^2(p^2+12)}}\right),
\label{eq3:15}\\
&u(x,t)=\varepsilon\omega\sqrt{\frac {3(p^2+4)}{\alpha\omega}}\left(
    {\frac {-36\beta\varepsilon\sqrt{-\frac {2\beta\omega}{p^2+4}}\wp^\prime(\xi)
    +12\beta{p}\left(3\beta\wp(\xi)+\omega\right)\wp(\xi)+\omega^2p}
    {36\beta^2p^2\wp^2(\xi)+12\beta\omega(p^2+6)\wp(\xi)+\omega^2(p^2+12)}}\right),
\label{eq3:16}
\end{align}
where $\wp(\xi)=\wp(x-{\omega}t,g_2,g_3)$, $\alpha\omega>0,\beta\omega<0$ and
the invariants $g_2$ and $g_3$ are given by (\ref{eq3:14}).\par
Because $\beta\omega < 0$ in (\ref{eq3:15}) and (\ref{eq3:16}), so
$\theta=-qp=-{\frac {2\omega}{\beta}}>0$. This indicates that the Weierstrass
elliptic function solutions (\ref{eq3:15}) and (\ref{eq3:16}) can only degenerate
to the solitary wave solutions of the mKdV equation. Therefore, by substituting
the conversion formulas (\ref{eq2:13}) and (\ref{eq2:14}) with $\theta=-qp=-{\frac {2\omega}{\beta}}$ into (\ref{eq3:15}) and (\ref{eq3:16}) we obtain the solitary
wave like solutions of the mKdV equation as follows
\begin{align*}
&u_1^{(4)}(x,t)=-\sqrt{\frac {3(p^2+4)}{\alpha\omega}}
   \left({\frac {4\sqrt{\frac {\omega^2}{p^2+4}}\sinh\xi\cosh\xi-\varepsilon\omega{p}}
   {4\cosh^2\xi)+p^2}}\right),\\
&u_2^{(4)}(x,t)=-\sqrt{\frac {3(p^2+4)}{\alpha\omega}}
   \left({\frac {4\sqrt{\frac {\omega^2}{p^2+4}}\sinh\xi\cosh\xi-\varepsilon\omega{p}}
   {4\sinh^2\xi-p^2}}\right),\\
&u_3^{(4)}(x,t)=\sqrt{\frac {3(p^2+4)}{\alpha\omega}}
   \left({\frac {4\sqrt{\frac {\omega^2}{p^2+4}}\sinh\xi\cosh\xi+\varepsilon\omega{p}}
   {4\cosh^2\xi)+p^2}}\right),\\
&u_4^{(4)}(x,t)=\sqrt{\frac {3(p^2+4)}{\alpha\omega}}
   \left({\frac {4\sqrt{\frac {\omega^2}{p^2+4}}\sinh\xi\cosh\xi+\varepsilon\omega{p}}
   {4\sinh^2\xi-p^2}}\right),\\
   &{\quad}\xi={\frac 12}\sqrt{-\frac {2\omega}{\beta}}(x-{\omega}t),\alpha\omega>0,\beta\omega<0,
\end{align*}
where $\omega$ is a free parameter.\par
\begin{figure}[htp]
\psfig{file=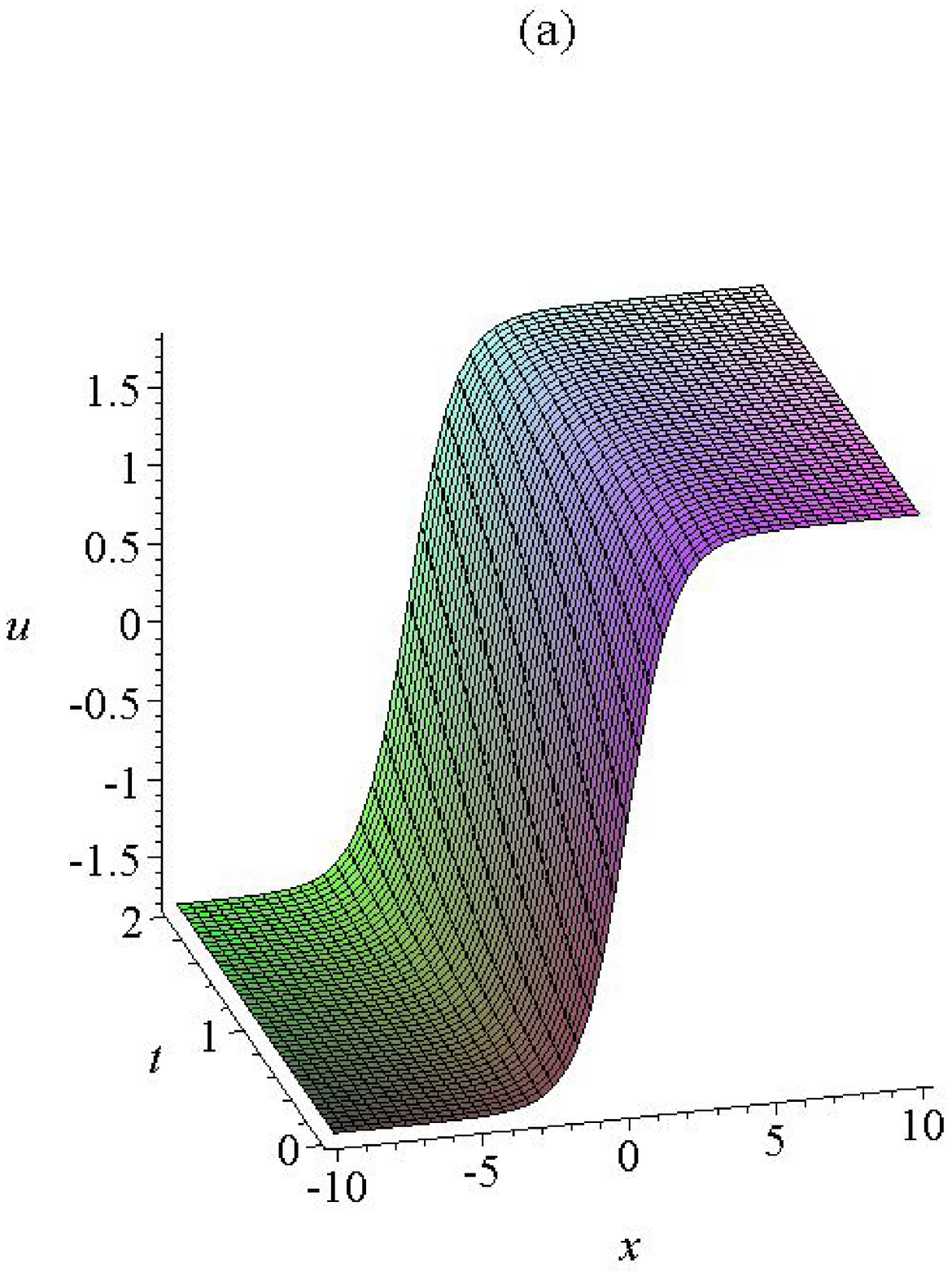,width=0.44\textwidth}
\psfig{file=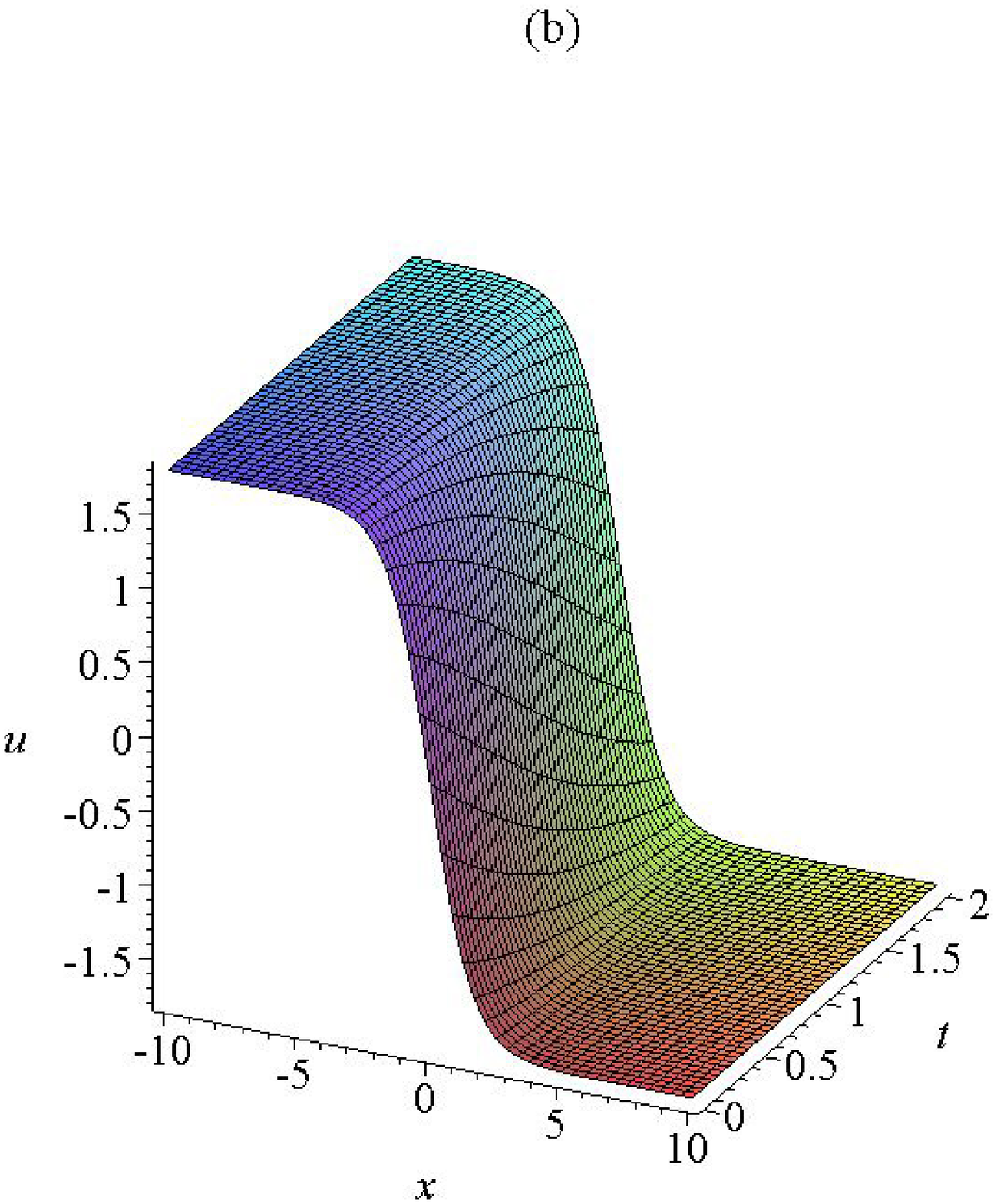,width=0.44\textwidth}
\psfig{file=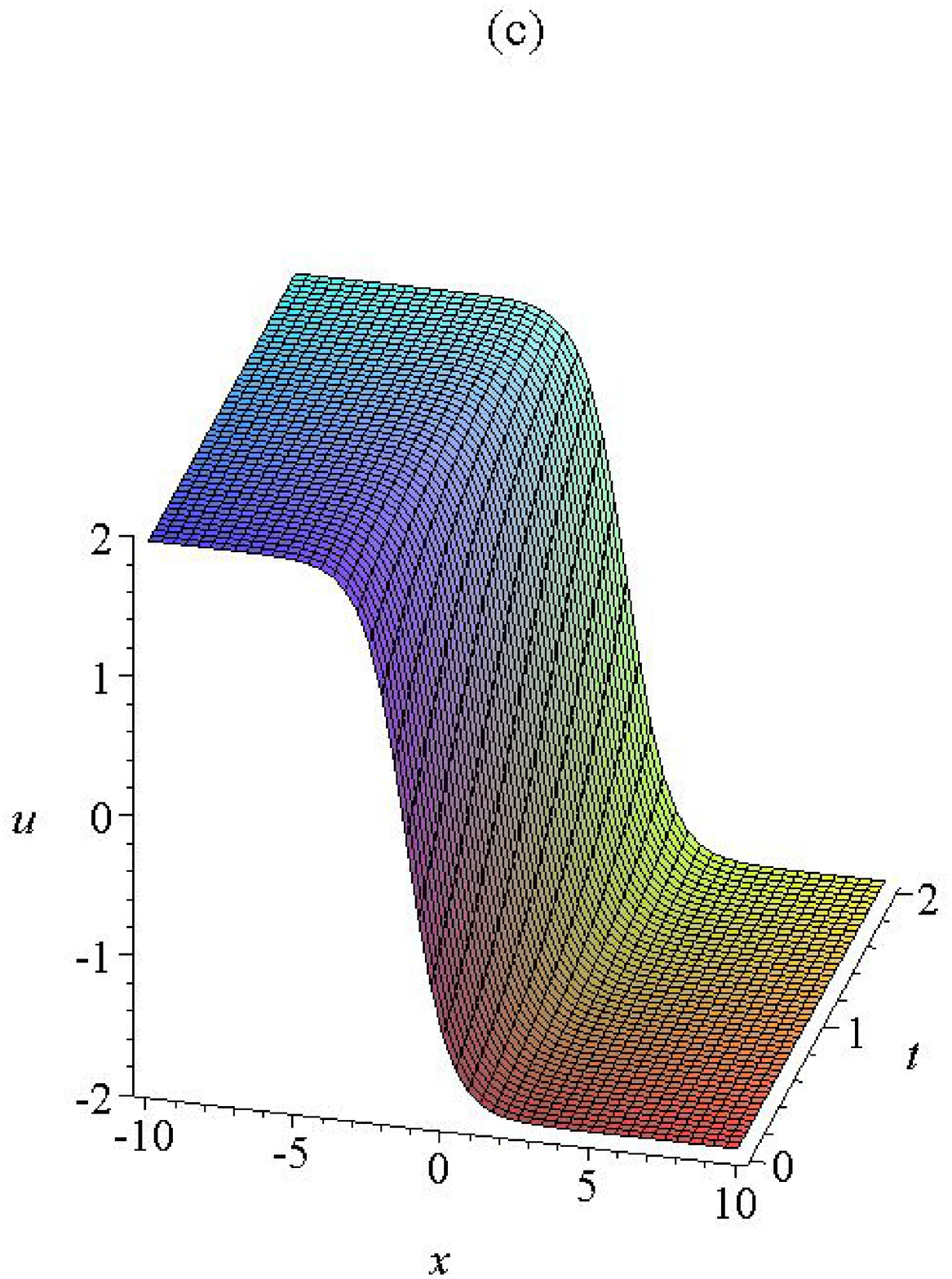,width=0.44\textwidth}
\hspace{28pt}
\psfig{file=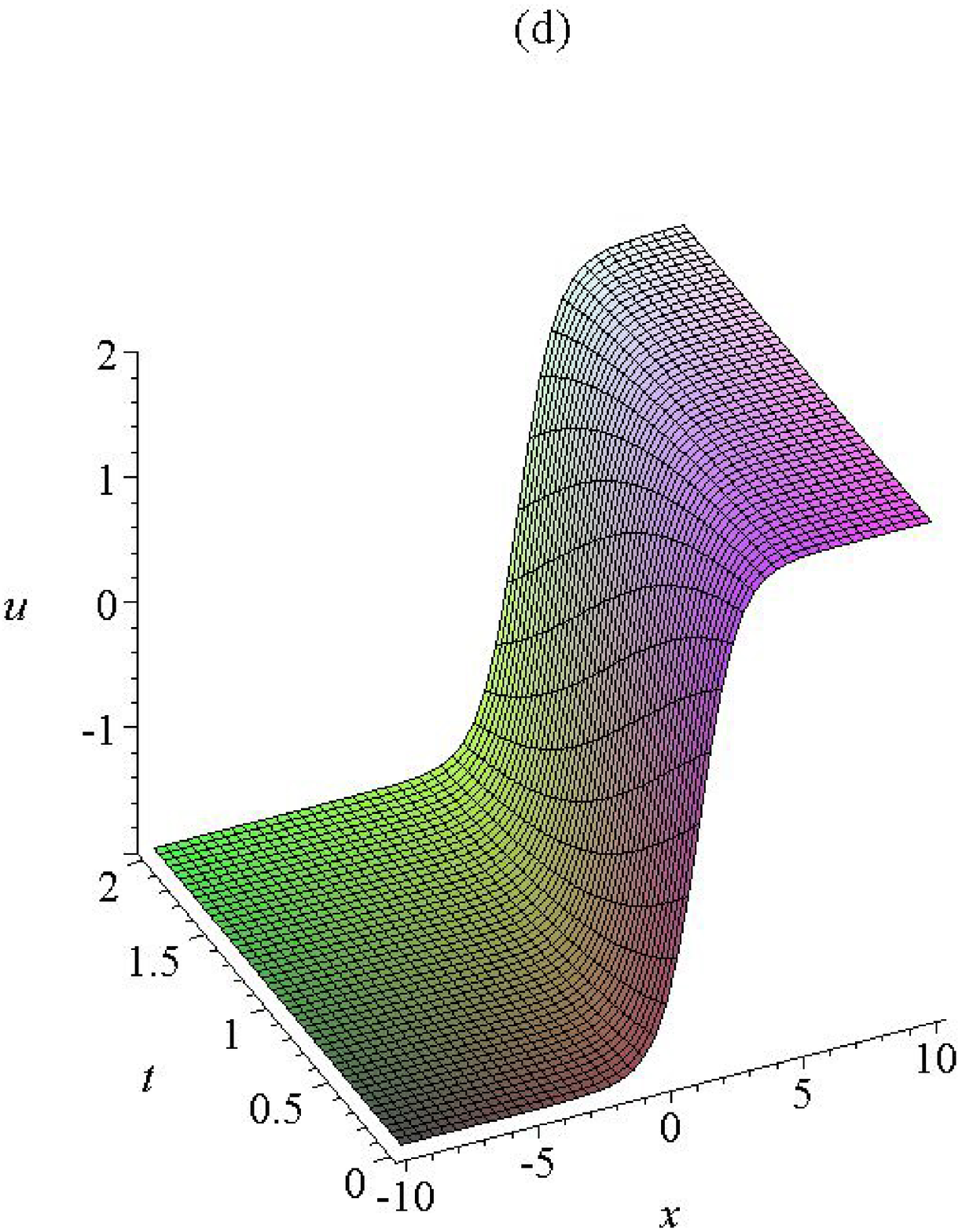,width=0.44\textwidth}
\vspace*{8pt}
\caption{The plots of the kink type and anti--kink type solitary wave solutions.
{\bf (a)} Kink type solution $u_1^{(1)}$ with $p=-1.25,q=1.25,\alpha=-1,\beta=1.35,\varepsilon=-1$,
{\bf (b)} Anti--kink type solution $u_1^{(1)}$ with $p=-1.25,q=1.25,\alpha=-1,\beta=1.35,\varepsilon=1$.
{\bf (c)} Anti--kink type solution $u_1^{(4)}$ with $p=2,\alpha=1,\beta=-1.25,\omega=1.25,\varepsilon=-1$,
{\bf (d)} Kink type solution $u_3^{(4)}$ with $p=2,\alpha=1,\beta=-1.25,\omega=1.25,\varepsilon=-1$.}
\label{fig:1}
\end{figure}
\begin{figure}[htp]
\psfig{file=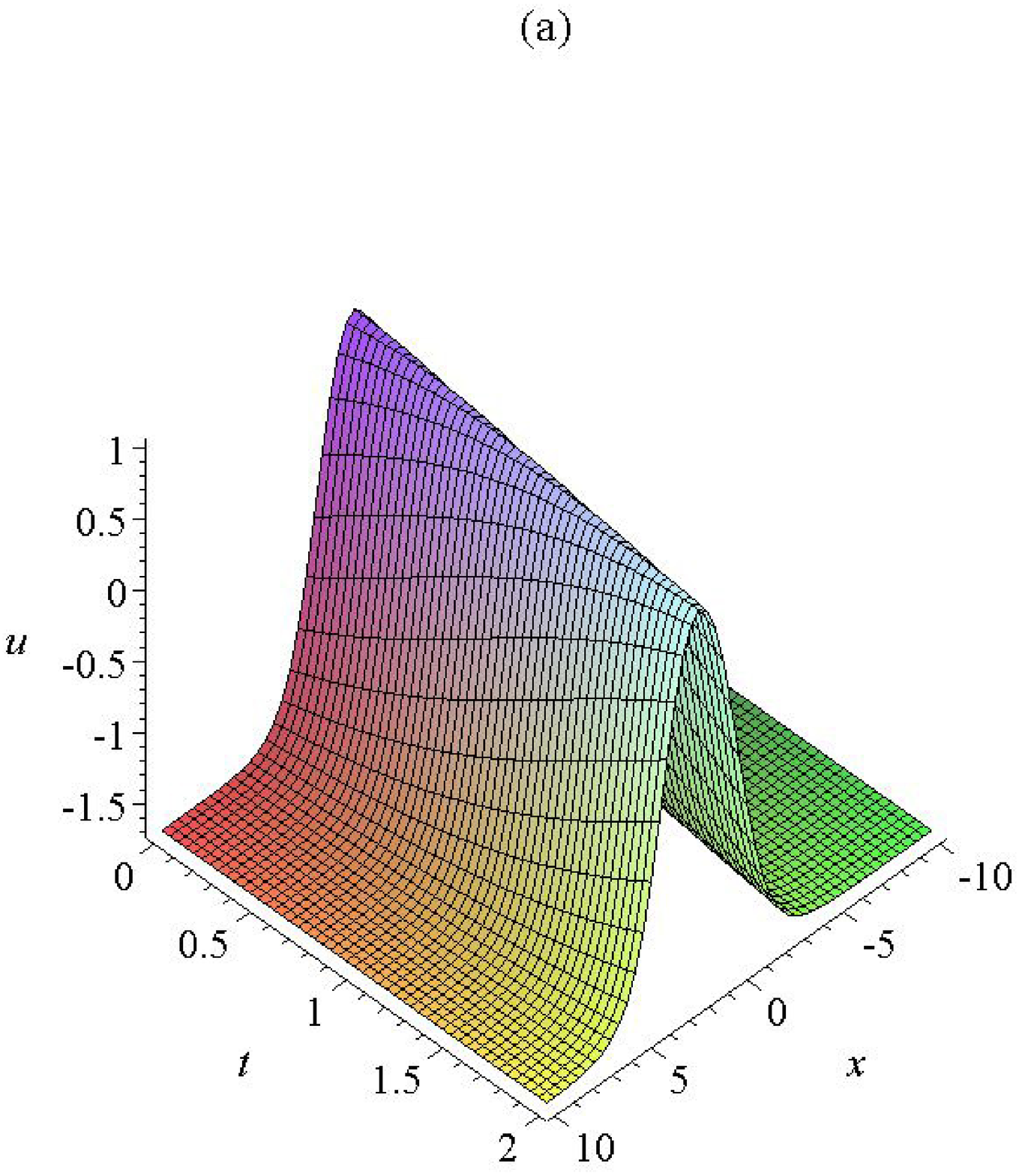,width=0.44\textwidth}
\psfig{file=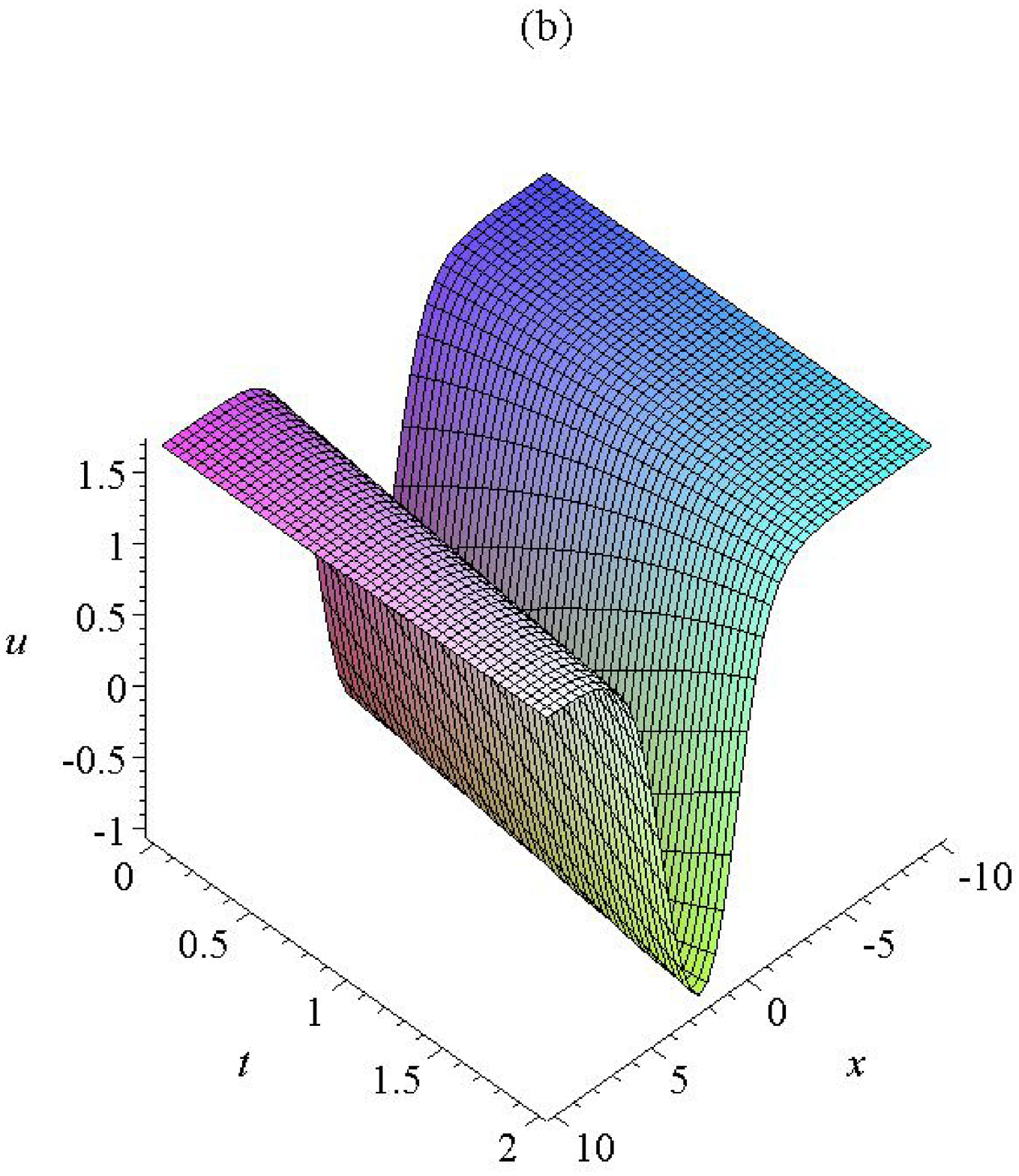,width=0.44\textwidth}
\psfig{file=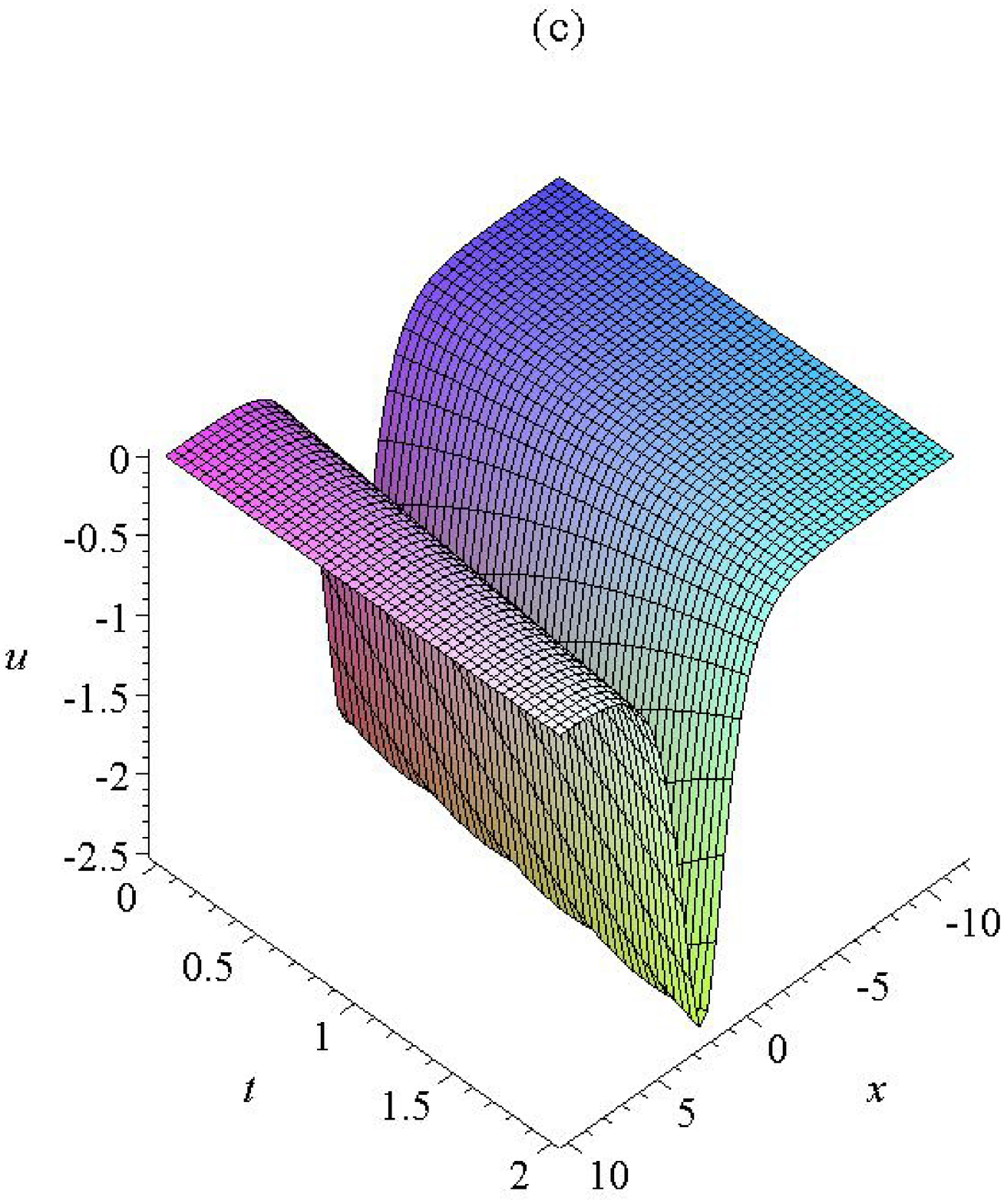,width=0.44\textwidth}
\psfig{file=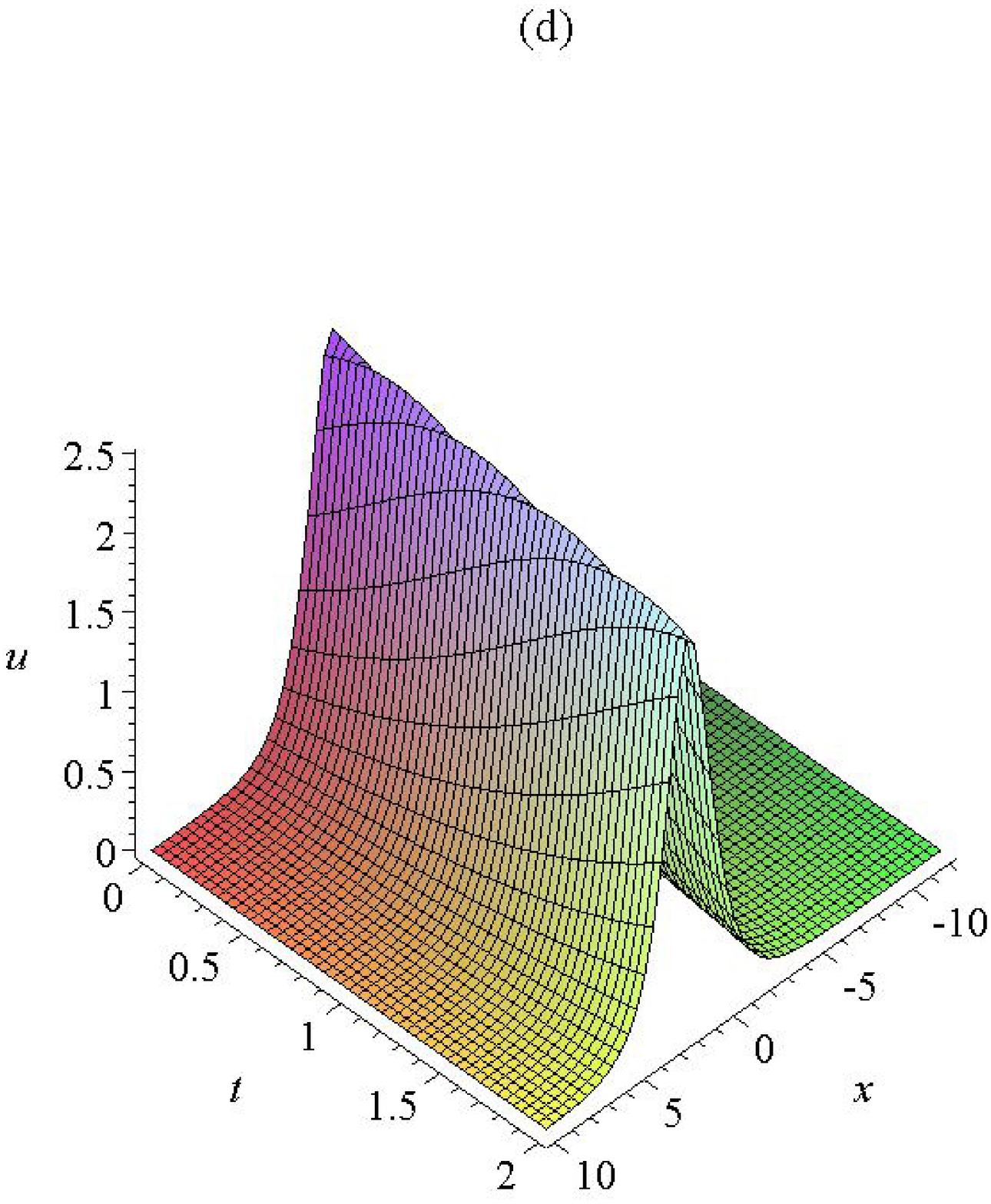,width=0.44\textwidth}
\psfig{file=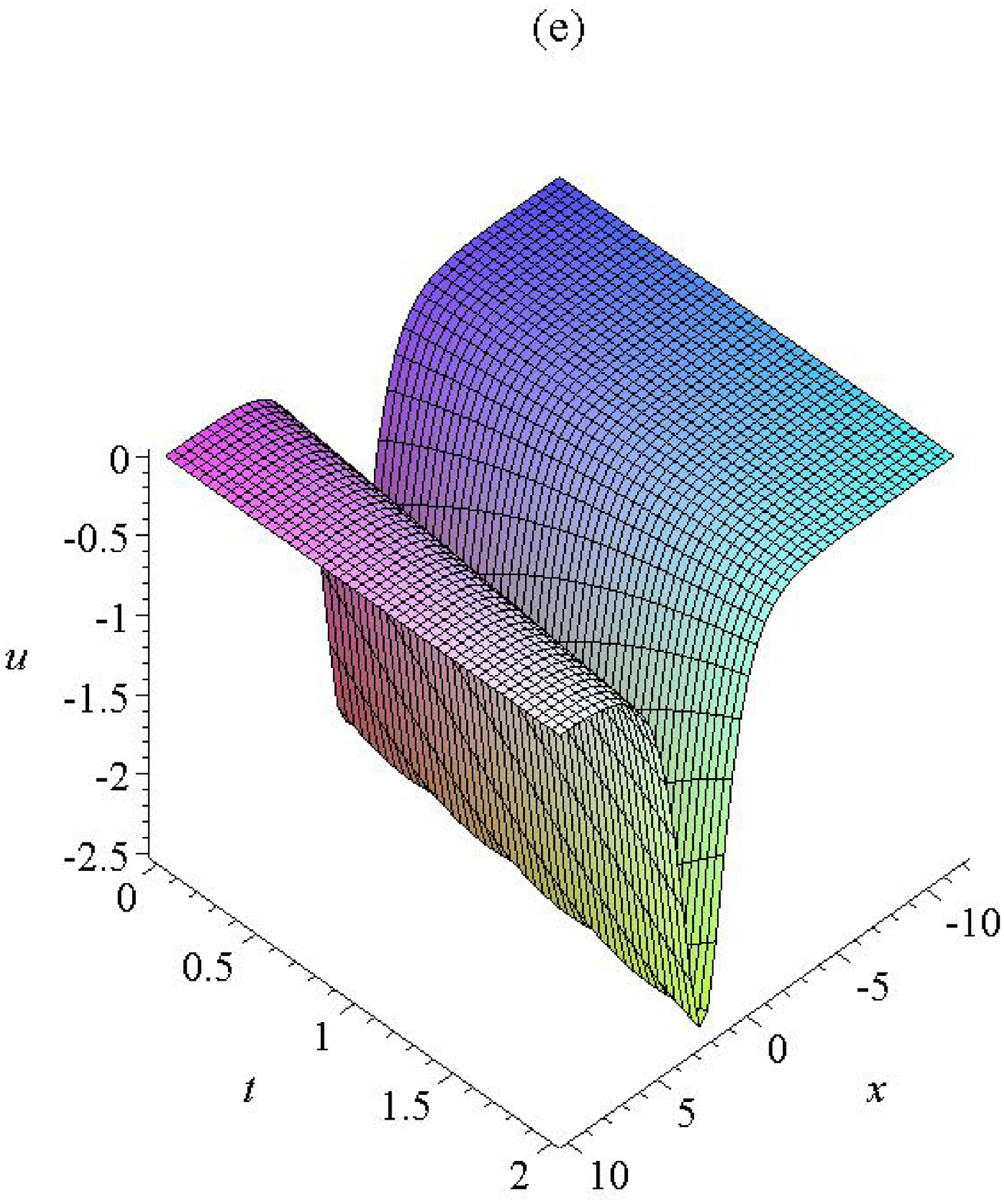,width=0.44\textwidth}
\hspace{34pt}
\psfig{file=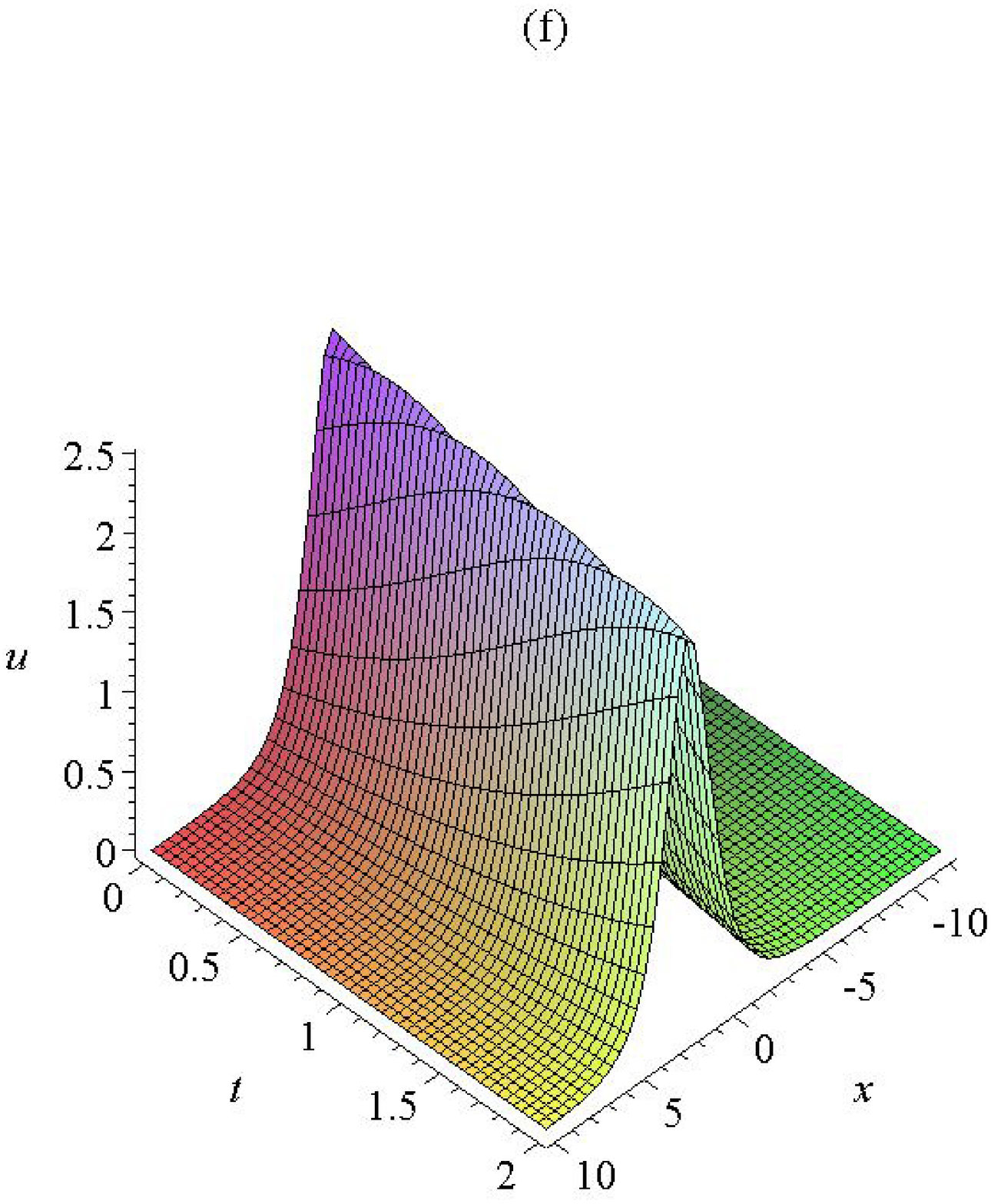,width=0.44\textwidth}
\vspace*{8pt}
\caption{The plots of the bell type and anti--bell type solitary wave like solutions.
{\bf (a)} Bell type solution $u_2^{(2)}$ with $\alpha=1,\beta=-1.25,\omega=1.025,\varepsilon=-1$,
{\bf (b)} Anti--bell type solution $u_2^{(2)}$ with $\alpha=1,\beta=-1.25,\omega=1.025,\varepsilon=1$.
{\bf (c)} Ant--bell type solution $u_1^{(3)}$ with $q=-2,\alpha=1,\beta=1.25,\omega=1.025,\varepsilon=-1$,
{\bf (d)} Bell type solution $u_1^{(3)}$ with $q=-2,\alpha=1,\beta=1.25,\omega=1.025,\varepsilon=1$.
{\bf (e)} Anti--bell type solution $u_2^{(3)}$ with $q=-2,\alpha=1,\beta=1.25,\omega=1.025,\varepsilon=-1$,
{\bf (f)} Bell type solution $u_2^{(3)}$ with $q=-2,\alpha=1,\beta=1.25,\omega=1.025,\varepsilon=1$.}
\label{fig:2}
\end{figure}
\begin{figure}[htp]
\psfig{file=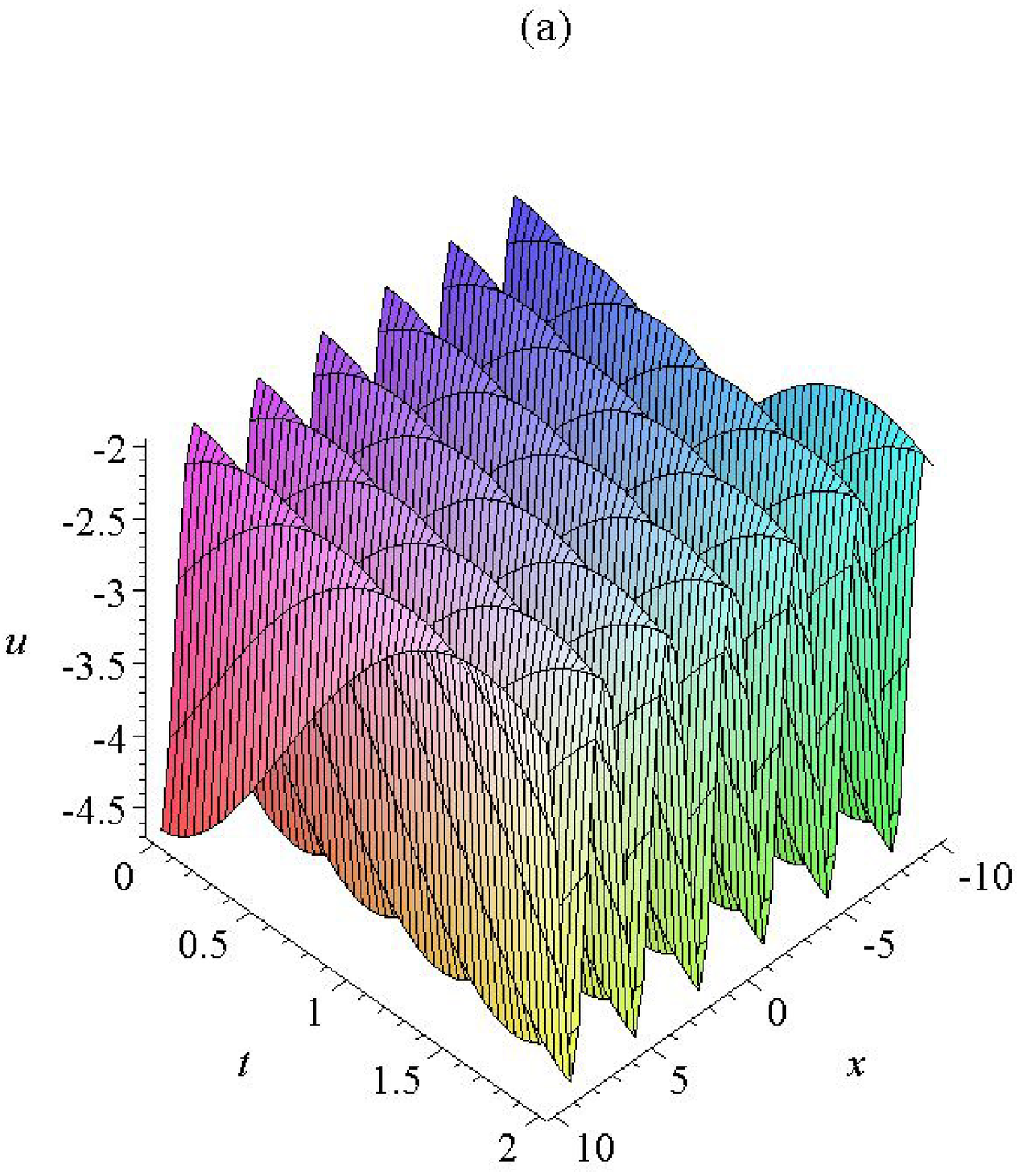,width=0.44\textwidth}
\psfig{file=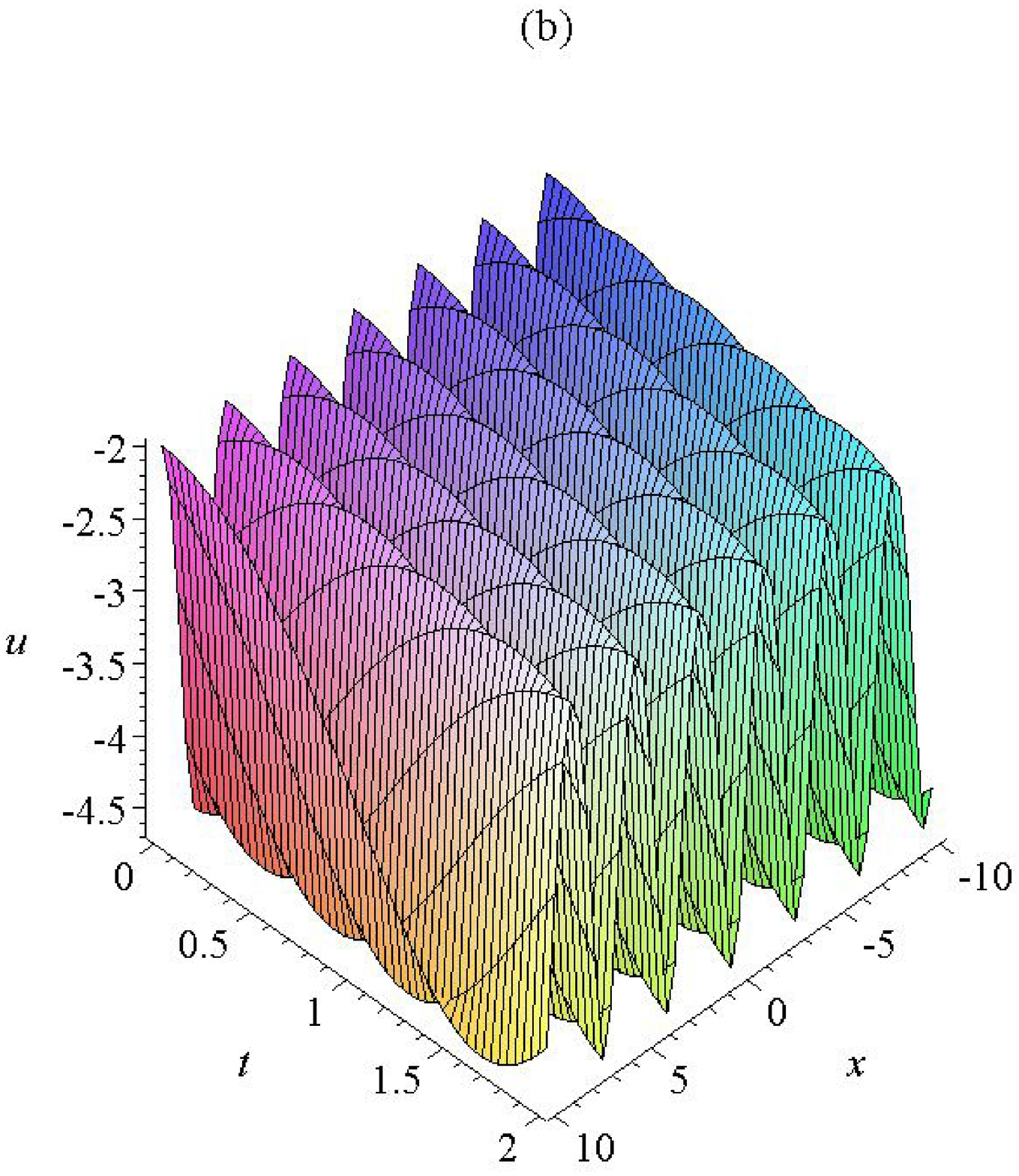,width=0.44\textwidth}
\psfig{file=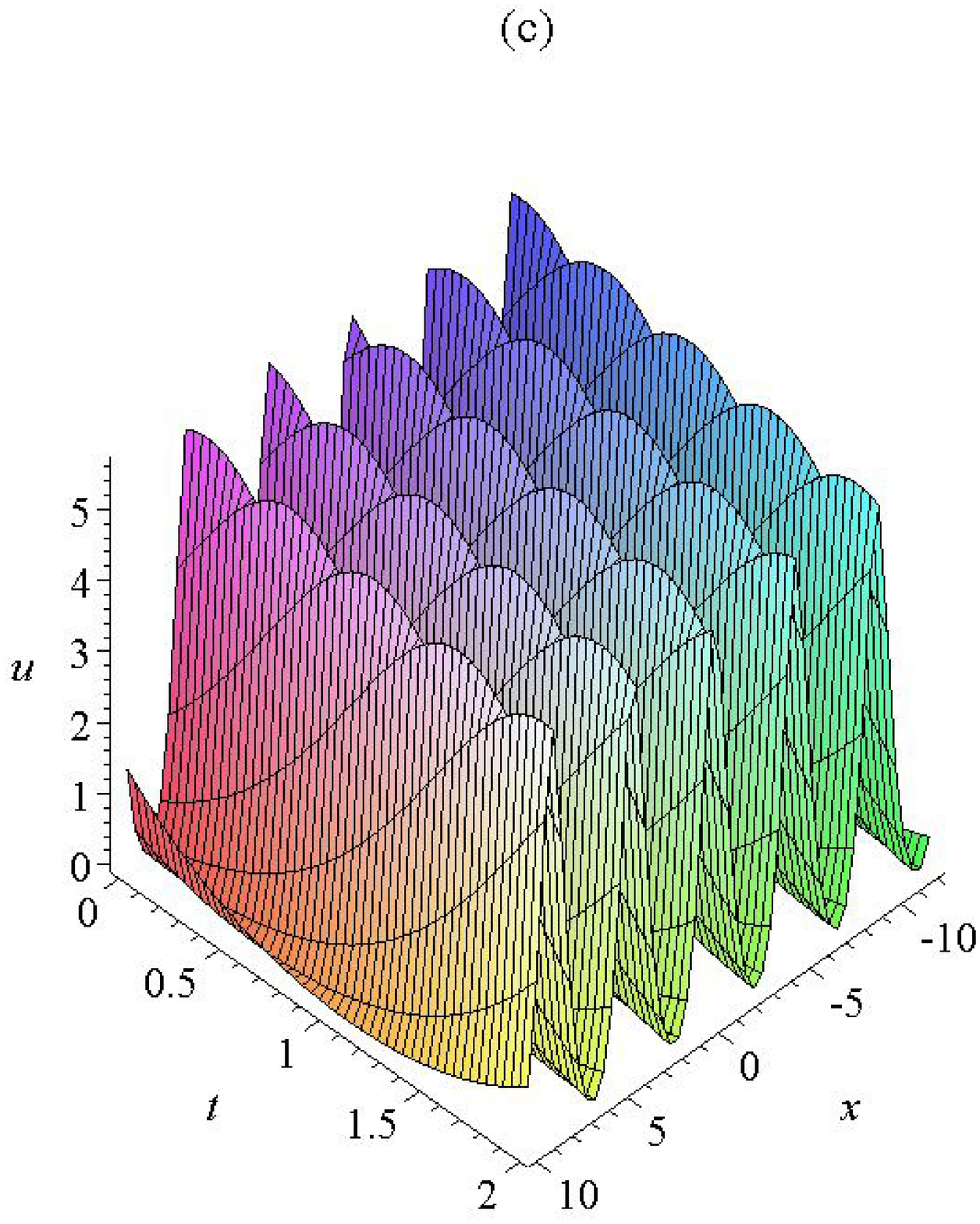,width=0.44\textwidth}
\hspace{36pt}
\psfig{file=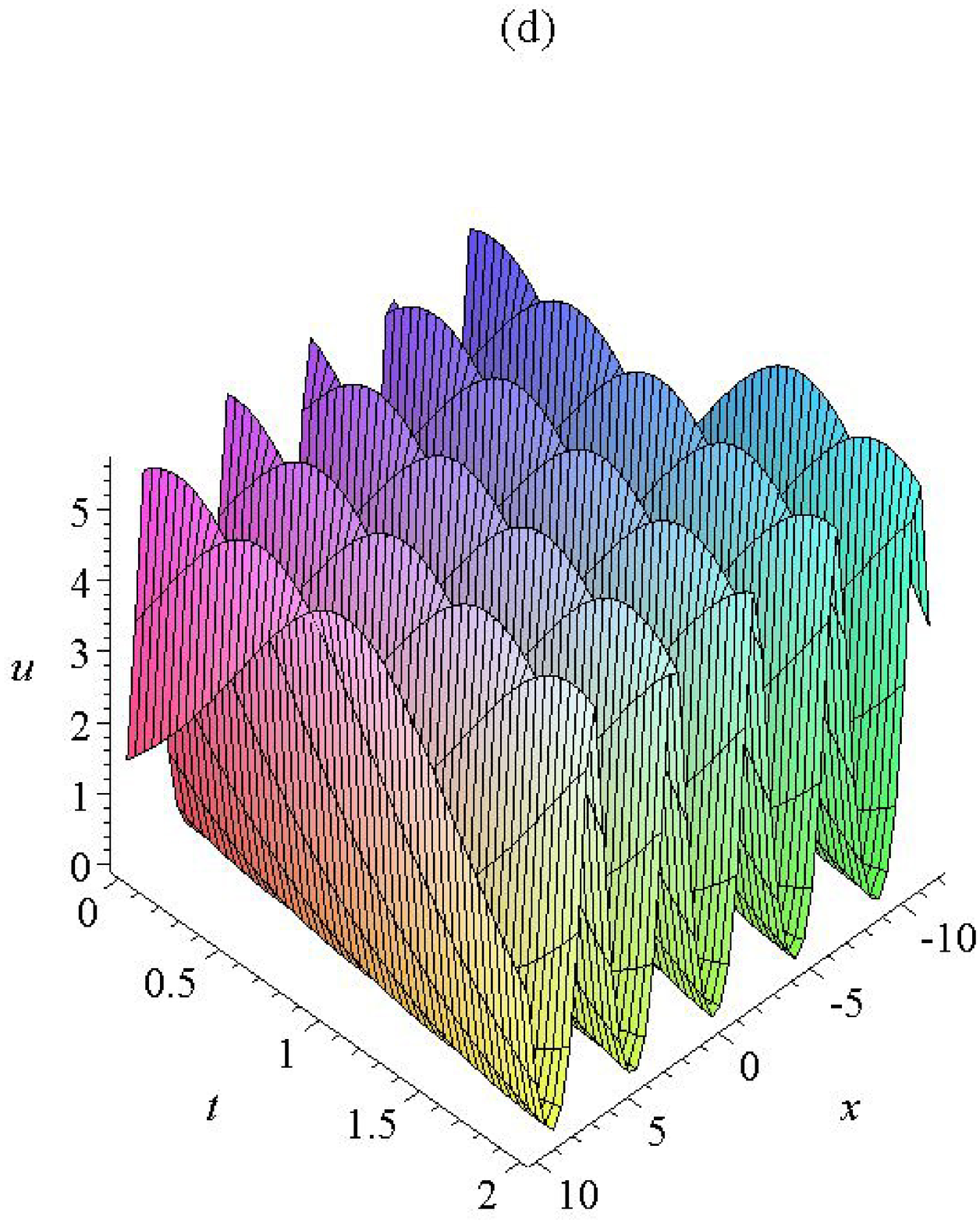,width=0.44\textwidth}
\vspace*{8pt}
\caption{The plots of the periodic solutions with $\alpha=0.25,\beta=0.5,\omega=0.995$.
{\bf (a)} Solution  $u_3^{(2)}$ with $\varepsilon=1$,
{\bf (b)} Solution $u_4^{(2)}$ with $\varepsilon=1$,
{\bf (c)} Solution $u_3^{(3)}$ with $q=2,\varepsilon=-1$,
{\bf (d)} Solution $u_4^{(3)}$ with $q=2,\varepsilon=-1$.
}
\label{fig:3}
\end{figure}

The above traveling wave solutions of the mKdV equation have not been obtained
in Refs.~\cite{17,18}, and the solutions $u_i^{(j)}~(i=1,2,3,4;j=2,3,4)$
cannot be obtained by using other direct methods.
More importantly, these solutions are very considerable in physics. For example,
the solutions $u_1^{(1)}$ and $u_2^{(2)}$ are the kink type and bell type solitary
waves for $\varepsilon=-1$, and for $\varepsilon=1$ they express the anti--kink
type and anti--bell type solitary waves, respectively.
The solutions $u_1^{(3)}$ and $u_2^{(3)}$ are the bell type and anti--bell type
solitary waves for $\varepsilon=1$ and $\varepsilon=-1$, respectively.
The solutions $u_1^{(4)}$ and  $u_3^{(4)}$ are the anti--kink type and kink type
solitary waves.
$u_3^{(i)},u_4^{(i)}~(i=2,3)$ are the trigonometric periodic solutions and other
solutions are singular solutions.
In order to observe the profiles of these solutions, the kink and anti--kink type
solutions, the bell and anti--bell type solutions and the periodic solutions are
shown in Fig.\ref{fig:1}, Fig.\ref{fig:2} and Fig.\ref{fig:3}. But the plots of
those singular solutions are omitted.

\section{Conclusions}
\label{sec-4}
In the present paper, the Weierstrass type projective Riccati equation
expansion method is proposed to construct Weierstrass elliptic function
solutions of NLEEs. At the same time, the conversion formulas are also
used to transform these Weierstrass elliptic function solutions into
the hyperbolic and trigonometric function solutions of NLEEs.
Our method is more powerful than the other direct algebraic methods, and
it can be regarded as an extension of the projective Riccati equation
expansion method. In order to explain our method more clear
we need to point out the following four points.\par
(1)\,The Weierstrass type projective Riccati equation expansion method
can give more types of new traveling wave solutions of NLEEs that cannot
be obtained by other direct methods. As shown above, in the case of the
mKdV equation, by using the Weierstrass type projective Riccati equation
expansion method we have obtained the kink type, the bell type, the
anti--kink type, the anti--bell type, the periodic and the singular
solitary wave and periodic solutions, etc. However, other direct methods
cannot give these types of solutions at the same time.
\par
(2)\,Compared with the previously known conversion formulas, our conversion
formulas (\ref{eq2:13})--(\ref{eq2:16}) don't require the roots of $p(w)=0$,
can convert the Weierstrass elliptic function solutions of NLEEs into the
hyperbolic and trigonometric function solutions in a straightforward way,
and can ensure that the obtained solutions are correct.\par
In addition, the conversion formulas (\ref{eq2:13})--(\ref{eq2:16}) also
can be used in other Weierstrass elliptic function methods to transform
the Weierstrass elliptic function solutions into the hyperbolic and trigonometric
function solutions.\par
(3)\,Although we have concerned with the mKdV equation, our method can be
applied to construct the exact solitary wave and periodic wave solutions
of a wide class of NLEEs.\par
(4)\,In addition to the projective Riccati equations, we find that other
auxiliary equations also possess the Weierstrass elliptic function solutions.\cite{21,24}
Therefore, the Weierstrass type expansion method can also be proposed for
these auxiliary equations which can lead us to establish a systematic
Weierstrass elliptic function method for solving NLEEs. So our idea is
significance to extend the application area of the Weierstrass elliptic
functions.

\section*{Acknowledgements}
This work is supported by the National Natural Science Foundation of China
under (Grant Nos. 11861050,11261037).

\end{document}